# Cryogenic avalanche detectors based on gas electron multipliers

A. Bondar, A. Buzulutskov [*], L. Shekhtman, R. Snopkov, Y. Tikhonov

*Budker Institute of Nuclear Physics, 630090 Novosibirsk, Russia*

**Abstract**

We study the performances of gaseous and two-phase (liquid-gas) cryogenic detectors of ionizing radiation based on gas electron multipliers (GEMs) and operated in an avalanche mode in pure noble gases. The gas amplification in He, Ar and Kr is systematically studied at low temperatures, using triple-GEM multipliers. High gains, exceeding $10^4$, were obtained in these gases in the range of 120-300 K. Stable electron avalanching was demonstrated in a saturated Kr vapor in the two-phase mode. These results are relevant for understanding basic mechanisms of electron avalanching at low temperatures and for applications in cryogenic particle detectors, in particular in dark matter and solar neutrino detectors.

*PACs:* 29.40.Cs; 34.80.My.

## 1. Introduction

Cryogenic detectors of ionizing radiation are relevant mostly in the field of noble liquid calorimetry, in high-energy physics experiments. Practically all such detectors are operated in an ionization mode, i.e. without internal amplification. On the other hand, for those cryogenic experiments where the primary ionization signal is weak, such as those of solar neutrino [1] and dark matter detection [2], it would be very attractive to operate in the electron-avalanching mode.

It should be remarked that the choice of the gas mixture for operation at low temperatures is limited due to the fact that most organic additives used in wire chambers are frozen out. Moreover, some of them should be avoided: it was reported that the wire chamber aging is dramatically enhanced at low temperatures in mixtures containing $CH_4$ [3]. Obviously, the operation with pure noble gases would solve the problem. Namely, the solution might be if one would obtain the electron avalanching either in the noble liquid itself, in single-phase detectors, or in the gas phase above the liquid, in two-phase detectors. However, the attempts

---

[*] Corresponding author. Email: buzulu@inp.nsk.su

to obtain high and stable avalanche amplification at low temperatures had a rather limited success.

Indeed, rather low gains (<100) were observed at low temperatures in liquid Xe [4,5] and Ar [6] and in gaseous Ar [7] using wire, needle or micro-strip proportional counters. Moreover, two-phase detectors [8-12], which initially seemed to solve the problem, turned out to have unstable operation in the avalanche mode: gain instabilities observed in two-phase Kr and Xe [9-12] were believed to arise due to condensation of a saturated vapor on wire electrodes.

The electron avalanching in dense noble gases (i.e. at low temperatures, in liquid phase, at high pressures) has a fundamental interest itself. For example, it was proposed that a new avalanche mechanism, namely the associative ionization (the Hornbeck-Molnar process [13]) dominates in light noble gases at high densities [14]. And the temperature dependence of avalanche processes might help to test this hypothesis. Another interesting feature is the possible formation of ion clusters at low temperatures [15]. Little is known however about electron avalanching at low temperatures. To our knowledge, there are only a few works dealing with this matter: in liquid Xe [4] and in gaseous He near 4 K [16].

We will show that the problem of electron avalanching at low temperatures, and consequently that of the development of cryogenic avalanche detectors, might be solved using Gas Electron Multipliers (GEMs) [17]. Indeed, it is known that the multi-GEM structures provide high gains ($\geq 10^4$) in all pure noble gases at room temperatures [18-21]. First evidences that the multi-GEM structures can successfully operate at low temperatures in gaseous and two-phase modes have been recently presented in our work [22]. In the present paper we elaborate on this subject. Namely, we present the detailed description of experimental procedures, describe the triple-GEM performance at low temperatures in gaseous and two-phase modes (in He, Ar and Kr), systematically study the electron avalanching as a function of temperature and give the interpretation of the results.

## 2. Experimental setup and procedure

The experimental setup is shown in Fig.1. The cryogenic avalanche detector is a cryostat with three GEM foils mounted in cascade inside. The cryostat consists of a vacuum-insulated chamber of a volume of 2.5 l, coupled to a flange with a number of high-voltage feedthroughs using an In seal. Two stainless steel windows, 0.1 mm thick and 10 mm diameter each, are made at the bottom of the chamber to transmit ionizing radiation. The cryostat is cooled using a heat exchanger mounted on the flange, filled with liquid nitrogen. The temperature inside is measured using a thermocouple, placed in the vicinity of the GEM assembly. The GEMs were produced by the CERN workshop. They have the following parameters: 50 μm thick Kapton, 70 and 55 μm hole diameter on the metal and kapton center respectively, 140 μm hole pitch, 28×28 mm$^2$ active area. The distances between the first GEM and the chamber bottom and between the GEMs are 5 mm and 2 mm, respectively. Electrical connections inside the chamber are provided using teflon-insulated wires.

The cryostat was operated at pressures reaching 3 atm either in the gaseous mode, in He, Ar and Kr, or in the two-phase (liquid-gas) mode, in Kr. It was filled with He, Ar or Kr of a purity higher than 99.99%. In addition, the oxygen impurity in Kr was reduced to a level of $2\times10^{-6}$ using a purification system of the liquid Kr calorimeter [23]. To further minimize impurities in He and Ar, these gases were passed through a liquid nitrogen trap when filling the chamber.

The detector was irradiated with an X-ray tube, having a molybdenum or rhenium target. The voltage and current supplied to the tube were in the range of 20-40 kV and below 60 µA, respectively. Typical X-ray fluxes (fluxes of absorbed photons) vary from $10^2$ to $10^4$ $s^{-1}mm^{-2}$. In the two-phase mode, the detector was also irradiated with β-particles from $^{90}$Sr source.

In the gaseous mode, either the copper foil (not shown in Fig.1), in He and Ar, or the bottom of the chamber made of stainless steel (SS), in He and Kr, acted as a cathode, thus providing two different cathode materials. This option is important for studying photon feedback effects. In the two-phase mode, only the bottom of the chamber acted as a cathode: an ionization produced in the liquid was extracted into the gas phase by an electric field, where it was detected with the help of the multi-GEM multiplier operated in a saturated vapor (Fig.1). The characteristic properties of operation in the gaseous and two-phase modes will be presented in appropriate sections.

The GEM electrodes were biased through a resistive high-voltage divider placed outside the cryostat, as shown in Fig.2. The characteristic property of the divider was that it consisted of three identical circuits connected in parallel, each GEM element being connected to one of them. Such a divider protects against discharges induced by ion feedback between GEM elements: even if one element breaks-down, the electrical potentials on other elements do not increase. This is not the case when using a single-circuit divider: we observed that all GEM elements could be destroyed after even a few discharges in Kr and Xe.

The anode signals were recorded from the last electrode of the third GEM (GEM3 in Fig.1), i.e. in a "3GEM" configuration according to terminology of Ref. [18], either in a current or pulse-counting mode. In the latter case, a charge-sensitive amplifier was used with a 10 ns rise time, 8 µs decay time and sensitivity of 0.5 V/pC.

Both in the gaseous and two-phase modes, the gain value was defined as the anode current, recorded from the third GEM, divided by the primary ionization current induced in the cathode gap. The latter current was determined in special measurements where the cathode gap was operated in ionization mode, the first GEM acting as an anode. The maximum attainable gain was defined as that at which no dark currents or discharges were observed for half a minute.

The basic idea of operation in the 3GEM configuration rather than in that of 3GEM+PCB, i.e. without a printed-circuit-board (PCB) electrode, is to be sensitive to signals induced by ions drifting through the GEM holes. In principal, this would allow to see whether the ion-induced signal in an avalanche depends on temperature, which may occur for example due to the formation of ion clusters. In order to be sensitive to such signals only and not to those induced by ions backdrifting between the third and second GEMs, the grounding capacitor (C1 in Fig.2) was connected to the first electrode of the third GEM.

## 3. Gaseous mode in He, Ar and Kr

### 3.1 Experimental procedure

In the gaseous mode the measurements were performed at a constant gas density: after the filling was completed, the gas supply was shut off. This allows to observe temperature effects induced by avalanche mechanisms other than the electron impact ionization (which is independent of temperature), in particular by those of atomic collisions. In addition, this allows to estimate the average temperature inside the chamber monitoring the pressure: for an ideal gas at a constant density the pressure is proportional to temperature. And all the gases studied can be considered as ideal with an accuracy of a few percents.

Cooling and measurement procedures in light (He) and heavy (Ar and Kr) noble gases were somewhat different. In He, the cooling and the measurements were carried out simultaneously.

Such a procedure benefits from additional purification of He, since most impurities would condense on the heat exchanger, at liquid nitrogen temperature.

In contrast, in Ar and Kr the measurements were carried out after the cooling procedure was completed (down to a given temperature), namely after the liquid nitrogen evaporated completely from the heat exchanger. This was to avoid Ar and Kr gas condensation at the heat exchanger during the measurements; otherwise the gas density would decrease to an uncertain value (the absence of condensate inside the chamber was verified monitoring pressure and temperature).

These procedures however may result in the appearance of some temperature gradients (up to 5 degrees) between the top and the bottom of the chamber, in particular induced by a heat inleakage from the flange. The temperature gradient results in some density gradient, which is presumably responsible for systematic errors shown in the following figures. The errors were determined comparing data from several measurement runs. The temperatures presented in the following figures, relevant to the gaseous mode, are those measured using the thermocouple, i.e. near the chamber bottom.

*3.2 Electron avalanching*

The electron avalanching at low temperatures was observed in all the noble gases studied. Fig.3 shows the triple-GEM gain as a function of the temperature in He, Ar and Kr at constant operation voltages and constant gas densities corresponding to pressures at room temperature of 3, 1 and 1 atm respectively.

One can see that the gain in He is independent of the temperature in the range of 120-300 K. The fact that the gain does not decrease when decreasing temperature rules out the effect of all organic and most inorganic impurities on the avalanche mechanism: they would be frozen out. That means that the unexpectedly high gains and ionization coefficients observed earlier in dense He at room temperatures [20,14] are not due to Penning ionization of impurities. Consequently, this indirectly supports the alternative explanation, namely the hypothesis of associative ionization [14].

In Ar and Kr the gain-temperature dependence is somewhat different from that of He (Fig.3): the gain first increases by a factor of 2-3 when decreasing temperature by 50 degrees and then goes onto plateau. The variation in gain of a factor of 2 corresponds to a considerably smaller variation in ionization coefficient, of about 10% at a gain of 1000, due to the fact that the gain is an exponential function of the ionization coefficient. Therefore the relative contribution of the mechanism responsible for the gain increase in Ar and Kr is small, of about 10%. This mechanism might be due to the atomic collision-induced ionization, in particular the associative ionization. It should be remarked that the cross-section of associative ionization could increase, decrease or be independent of temperature, depending on atomic potentials and collision energies, similarly to Penning ionization [24,25]. Unfortunately, theoretical predictions for He, Ar and Kr are not available, so that the associative ionization hypothesis cannot be tested at the moment using temperature dependence.

Gain-voltage characteristics of the triple-GEM at room and low temperatures in gaseous He and Ar are shown in Fig.4 and in gaseous Kr in Fig.5(top). In He, the characteristic is shown at 123 K only, since it coincides with that of room temperature. In He and Ar, the gas densities correspond to pressures at room temperature of 3 and 1 atm respectively. In Kr, the data sets for two different gas densities are presented corresponding to pressures at room temperature of 1 and 2.5 atm. The latter density is equal to the density of saturated Kr vapor in the two-phase mode at 1 atm; a comparison between the gaseous and two-phase modes will be done in section 4.3.

One can see that rather high gains are reached in all gases at low temperatures. In particular, the maximum gain exceeds $10^5$ and a few tens of thousands in He and Ar and Kr, respectively. Note that the slopes of the gain curves in Ar and Kr slightly increase at low temperatures, thus suggesting a small increase of ionization coefficients as discussed above.

*3.3 Charging-up effect*

One of the main objectives of the current study was to check if GEMs could operate at low temperatures. It is known that GEMs may undergo charging-up effects at high incident fluxes and high anode current densities, induced by positive ion deposition on a kapton surface [26,27,18]. The charging-up effect manifests itself in that the gain-voltage characteristic depends on the flux: the gain increase with voltage is faster for higher fluxes. And it might be possible that the charging-up would be so strong at low temperatures due to an enhanced resistivity of kapton that the GEM would not be able to operate even at very low incident fluxes.

Indeed, the resistance of a GEM foil increases by an order of magnitude per 30 degrees as seen from Fig.6: the leakage current across a single GEM is shown as a function of the reciprocal temperature, in He and Ar. Direct calculations using the surface resistivity of kapton (about $10^{16}$ Ohm per square at room temperature [28]) indicate that the leakage current can be entirely dominated by surface current in the GEM holes, while the contribution due to the bulk conduction (with kapton resistivity of $10^{17}$ Ohm·cm [28]) can be neglected.

The temperature dependence of the conductivity, bulk or surface, obeys the Boltzman-type law:

$\sigma = \sigma_0 \exp(-E_A / kT)$.

Here $E_A$ is the activation energy of the conduction. Extrapolating this dependence in Fig.6, the kapton surface conductivity would decrease considerably, by 6 orders of magnitude, when decreasing temperature from 295 to 120 K.

Nevertheless, we did not observe any indication on charging-up effects at low temperatures, as well as at room temperatures. In particular, Fig.7 shows gain-voltage characteristics of the triple-GEM in He at 144 K at different primary ionization fluxes. One can see that there is no dependence on the flux, despite the fact that the flux value varies by 2 orders of magnitude, from $9 \times 10^3$ to $8 \times 10^5$ electron/(mm$^2$s). In Ar, the gain was also independent of the flux at low temperatures. These observations demonstrate that GEMs can operate at low temperatures at high gains and high fluxes regardless of the kapton resistivity, probably even down to liquid He temperatures.

*3.4 Pulse shape and photon feedback*

Figs.8-10 show typical anode signals from the triple-GEM in He, Ar and Kr, using either the Cu or stainless steel cathode. Gain values indicated in the figures are those estimated in the current mode. In general, we do not observe any unusual properties in the shape of anode pulses induced just by low temperatures. This is seen when comparing Fig.8(middle) to Fig.8(bottom), showing the anode signals in He at 295 and 124 K respectively, using the stainless steel cathode. In both cases the anode pulse has a characteristic triangular shape: the linear pulse rise corresponds to integration of the primary signal induced by electrons and ions drifting through the holes of the third GEM. This was also valid for signals in Kr at all temperatures (Fig.10,top) and in Ar at room temperature (Fig.9,top). The width of the primary signal in all gases is independent of the temperatures: it is in the range of 100-200 ns (FWHM).

Fig.10(top) shows a sample of signals obtained in Kr using the X-ray tube with a Mo target, providing characteristic X-ray lines around 18 keV. One can see here two groups of pulses: the group with smaller pulses obviously corresponds to the "escape peak". The rather high gain obtained in Kr at 180 K, of about $1.8\times10^4$, should be emphasized.

We did not observe any indication of photon feedback using the stainless steel cathode. On the other hand, using the Cu cathode the photon feedback effect was observed in He at all temperatures, in accordance with earlier observations [21]: at high gains, exceeding few thousands, the primary signal is accompanied by a secondary signal (Fig.8,top). Detailed analysis of the data of the present work and Ref. [21] leads to a conclusion that secondary signals are induced mainly by photon feedback between the last GEM element and the cathode. The photon feedback results also in faster gain increase with voltage at high gains: one can see this comparing data in Fig.4 obtained using the stainless steel cathode to those in Fig.7 obtained using the Cu cathode.

The photon feedback in He is explained by a relatively high value (of about 10%) of the quantum efficiency of the Cu photocathode in the emission region of He (80 nm) [29]. On the other hand, the quantum efficiency of Cu is considerably reduced at the emission region of Ar (130 nm) and Kr (150 nm). Therefore the anode signal in Ar and Kr has no secondary signals at room temperatures, even using the Cu cathode (see Fig.9,top and Ref. [21]). An interesting observation is that the photon feedback in Ar appears at 170 K as seen from Fig.9(bottom). One of the possible explanations might be the enhancement of the quantum efficiency of Cu photocathodes at low temperatures, similarly to that observed for CsI photocathodes in liquid Kr and Xe [30].

## 4. Two-phase mode in Kr

*4.1 Experimental procedure*

The characteristic of cooling and measurement procedures in the two-phase mode was that the cryostat was connected to the gas supply at all times, namely to a 40 l bottle of Kr at an initial pressure of 1.9 atm. Such an amount of Kr in the system provides a 3 mm thick liquid layer at the end of the cooling procedure, at a temperature of 120 K; such a layer thickness is enough for full absorption of X-rays and β-particles.

Since there were no level meters in the setup, we used some test procedures to be convinced of the formation of the liquid phase. It is known that the electron emission from a liquid into the gas phase has a threshold behavior due to a potential barrier at the liquid-gas interface [10-12]: it takes place only if the electric field exceeds a critical value, of about 2 kV/cm in Kr. Therefore, the signature for the liquid layer formation should be a disappearance of the anode current recorded in the cathode gap (first GEM acting as an anode) at a field value close or below the threshold.

This is illustrated in Fig.11 showing the time evolution, during the cooling cycle, of the following quantities: the anode current in the cathode gap at a field of 1.9 kV/cm induced by X-rays, the thermocouple temperature and the pressure-to-temperature ratio. One can see, that the current first increases due to increasing X-ray absorption: this is because the gas density increases when decreasing temperature. At some point, it stops increasing, corresponding to full X-ray absorption, and then rapidly drops, indicating the liquid phase formation.

In addition, the *p/T* curve in Fig.11 shows a kink at the time corresponding to liquid phase formation. This is because the dependence between the vapor pressure and the temperature is much stronger in two-phase systems than in single phase. Due to this fact, the pressure monitoring in the two-phase mode gives a more precise temperature estimation than using the thermocouple. Accordingly, the temperatures presented in the following figures, relevant to the two-phase mode, are those of estimated from pressure monitoring.

*4.2 Electron collection in the cathode gap*

The electron emission from liquid Kr as a function of the electric field is illustrated in

Fig.12. In the gaseous mode the anode current recorded in the cathode gap is independent of the field. In contrast, in the two-phase mode the current is recorded only when the electric field exceeds 2 kV/cm, as discussed above. Therefore, during the measurements in the two-phase mode, the electric field in the cathode gap was kept above the critical value, in the range of 2.5-3.5 kV/cm. To this extent, the last resistance in the voltage divider, defining the electric field in the cathode gap, was increased by a factor of 3.5 as compared to the gaseous mode (Fig.2).

Let us estimate the collection time of electrons from liquid Kr. The electron drift velocity in liquid Kr at 3 kV/cm is $3\times10^5$ cm/s, and the absorption length of a 20 keV photon and the range of a 1 MeV electron are 75 µm and 1.9 mm respectively. Consequently, the average drift times of electrons produced by X-rays and β-particles in a 3 mm thick liquid layer are estimated to be 1 and 0.7 µs, respectively.

Now we can estimate the electron collection efficiency from liquid Kr. For the given impurity level of oxygen of $2\times10^{-6}$, the electron life-time in liquid Kr is 0.4 µs [12]. Therefore, the primary ionization signal in liquid Kr would be reduced, due to the electron capture, by a factor of 12 and 6 for X-rays and β-particles, respectively. The signal would be further reduced due to the fact that the electron emission probability from liquid Kr is about 60% at 3 kV/cm [10,12]. In addition, the electron-ion recombination in the liquid may also reduce the signal, by a factor of 1.1-1.2 [12,31]. Combined effect of these factors results in that the electron collection efficiency in the cathode gap in the two-phase mode is by a factor of 20 and 10 lower than that in the gaseous mode, for X-rays and β-particles respectively. This is seen in Fig.12 for X-ray-induced signals.

*4.3 Electron avalanching*

We have observed the electron avalanching in the two-phase mode in Kr, in a saturated vapor. The detector operated for at least an hour in the two-phase mode without visible degradation of the gain. The appropriate gain-voltage characteristics of the triple-GEM are shown in Fig.5(bottom). Two data sets are presented, at saturated vapor pressures of 0.84 and 1.32 atm corresponding to temperatures of 118 and 123 K. At both temperatures the gain reaches $10^4$. At the maximum gain, some secondary processes seem to start playing a role, since the gain increase with voltage becomes too fast.

The gaseous and two-phase data are in reasonable agreement. At equal gas densities, the operation voltages and the maximum gains in the two-phase mode are generally similar to those of the gaseous mode: this is seen comparing data in Fig.5(top) and Fig.5(bottom). This means that the electron avalanching in the saturated vapor does not differ from that of the normal gas, in general.

Fig.10(bottom) further illustrates the detector performance in the two-phase mode: anode signals of the triple-GEM are shown at a vapor pressure of 0.94 atm, corresponding to a temperature of 119 K. The signals are induced by β-particles of $^{90}$Sr. From this figure one can deduce the collection efficiency of electrons created in liquid Kr by β-particles. Taking into account the average pulse-height ($A$=0.6 V), amplifier calibration ($k$=0.5 V/pC), triple-GEM gain ($G$=900), average energy deposited in the liquid by β-particles ($E$=1 MeV), energy needed for ion pair creation in liquid Kr ($W$=20 eV) and using the formula
$A/k = \varepsilon\, G\, E/W$,
the collection efficiency turns out to be $\varepsilon$=17%. This is not far from the value calculated in section 4.2 (10%).

Comparing Fig.10(top) and Fig.10(bottom), one may conclude that the pulse shape in the two-phase mode does not differ from that of the gaseous mode. One should notice however that the pulse-height in the two-phase mode is larger than that in the gaseous mode, despite the fact that the gain is by a factor of 20 lower. This is because the energy deposited by β-particles in liquid Kr (~1 MeV) is much larger than that deposited by X-rays in gaseous Kr (~20 keV).

It is interesting to estimate the Townsend ionization coefficients from gain-voltage characteristics and to compare them at different temperatures and gas densities. Here we follow the approach described in Ref. [14]: the GEM hole is approximated by a parallel-plate counter with an electric field taken equal to that calculated in the center of the hole. The result is presented in Fig.13: the reduced ionization coefficient (i.e. normalized to atomic density) is shown as a function of the reduced electric field. The important observation is the scaling behavior of ionization coefficients obtained in the two-phase mode at different pressures: this confirms that the data presented in Fig.5 are consistent. Also, the data obtained at room temperature are in reasonable agreement with those taken from literature obtained at low pressures [32]. On the other hand, the ionization coefficients obtained in the two-phase mode at 118 and 123 K are somewhat larger than at room temperature, by about a factor of 1.5. This fact supports the conclusion made in section 3.2 that the avalanche mechanism in Kr is somewhat modified at low temperatures.

In section 4.2 we demonstrated the existence of the liquid phase in an ionization mode (Fig.12). To demonstrate this in an avalanche mode, special-purpose measurements were carried out. The idea is based on the fact that a considerably larger energy is deposited by β-particles in the liquid compared to that of the gas. Accordingly, the primary ionization current in the cathode gap in the two-phase mode was compared to that of the gaseous mode. This current is equal to the ratio of the anode current of the triple-GEM to its gain: $I_C=I_A/G$. Fig.14 shows this ratio as a function of the pressure. Gain values in the two-phase mode, for given voltages and gas densities, were calculated from Fig.13 using the fitted (dashed) curve. The gain variations are rather large: in the range of $7\times10^2$-$1.5\times10^4$. Nevertheless, two-phase data points in Fig.14 are grouped together and the difference between the two-phase and gaseous modes is distinctly seen: it is of a factor of 10.

One can notice, that systematic errors in Figs.5,12-14 in the two-phase mode are always larger than in the gaseous mode. This is due to rather large fluctuations of primary ionization current observed in the two-phase mode (Fig.12). We suppose that these fluctuations were induced by surface waves in the liquid, generated by drops of condensed Kr falling from the heat exchanger. Indeed, in the measurements where the liquid nitrogen in the heat exchanger was replaced with cold gaseous nitrogen, thus providing lower condensation rate of Kr, the fluctuations were reduced. This is seen in Fig.5(bottom), where the data set with larger errors was obtained with liquid nitrogen in the heat exchanger, while that with smaller errors with cold gaseous nitrogen. This also means that in the future design of two-phase detectors one should avoid the presence of the surface waves.

## 5. Conclusions

We have studied the performances of gaseous and two-phase (liquid-gas) cryogenic detectors of ionizing radiation based on gas electron multipliers (GEMs) and operated in an avalanche mode in pure noble gases. We confirmed the results obtained in Ref. [22].

Using triple-GEM multipliers, the electron avalanching was systematically studied in He, Ar and Kr in the range of 120-300 K. High gains, exceeding $10^4$, were obtained in these gases at low temperatures. In He, the electron avalanching is independent of the temperature. In Ar and Kr, it has moderate temperature dependence: the gain increases by a factor of 1-5 when decreasing temperature.

The performances of triple-GEM structures in the two-phase mode in Kr were investigated. In particular, a stable electron avalanching was demonstrated in a saturated vapor in Kr; gain values, exceeding $10^3$, were obtained. In general, the electron avalanching in a saturated vapor does not differ from that of the normal gas, in terms of gain-voltage characteristics and pulse-shapes.

No charging-up effects were observed in He and Ar indicating that the GEM structures can successfully operate at low temperatures, probably even down to liquid He temperatures. Using a Cu cathode, photon feedback effects were observed in He and Ar.

The results obtained are relevant for understanding basic mechanisms of electron avalanching at low temperatures and for applications in cryogenic particle detectors, in particular for dark matter and solar neutrino detectors. Further studies of this technique are on the way.

**Acknowledgements**

The research described in this publication was made possible in part by Award No. 2550 of the U.S. Civilian Research & Development Foundation for the Independent States of the Former Soviet Union (CRDF). This work has been partially motivated by the possible application in cryogenic two-phase detectors for solar neutrino and dark matter search. We are indebted to Prof. W. Willis and Drs. J. Dodd and M. Leltchouk, of the Columbia University, and Dr. D. Tovey, of the Sheffield University, for having suggested these applications.

25. B. M. Smirnov, Excited atoms, Energoizdat, Moscow, 1982 (in Russian).
26. R. Bouclier, W. Dominik, M. Hoch, J. C. Labbe, G. Million, L. Ropelewski, F. Sauli, A. Sharma, G. Manzin, Nucl. Instr. and Meth. A 396 (1997) 50.
27. A. Bressan, A. Buzulutskov, L. Ropelewski, F. Sauli, L. Shekhtman, Nucl. Instr. Meth. A 423 (1999) 119.
28. DuPont High Performance Materials, http://www.dupont.com/kapton.
29. B. B. Cairns, J. A. R. Samson, J. Opt. Soc. Amer. 56 (1966) 1568.
30. E. Aprile, A. Bolotnikov, D. Chen, F. Xu, V. Peskov, Nucl. Instr. and Meth. A 353 (1994) 55.
31. S. Kubota, M. Hishida, M. Suzuki, J. Ruan, Phys. Rev. B 20 (1980) 3486.
32. Yu. P. Raizer, Gas Discharge Physics, Springer, Berlin, 1997.

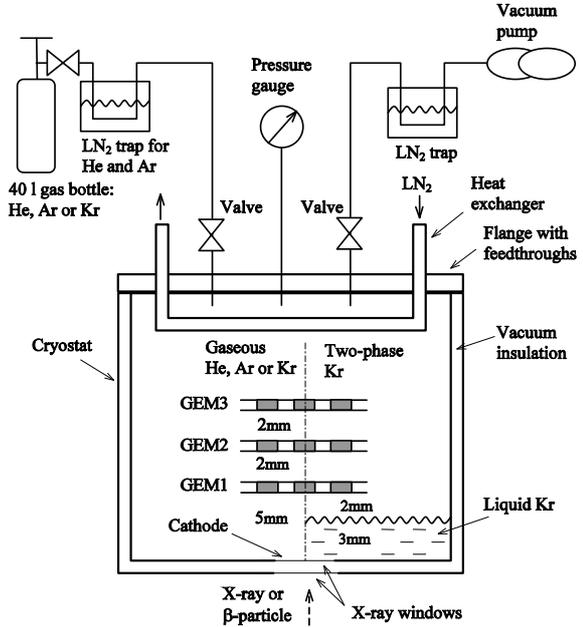

Fig.1 Schematic view of the cryogenic avalanche detector operated in the gaseous and two-phase modes in He, Ar and Kr.

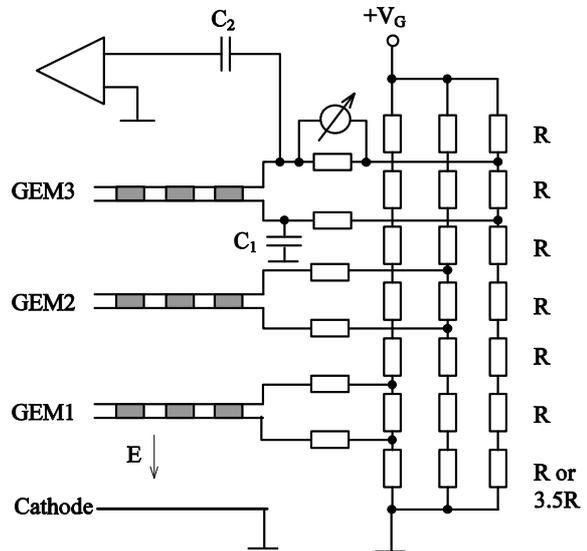

Fig.2 Electrical connections of the triple-GEM to a high-voltage divider and readout electronics.

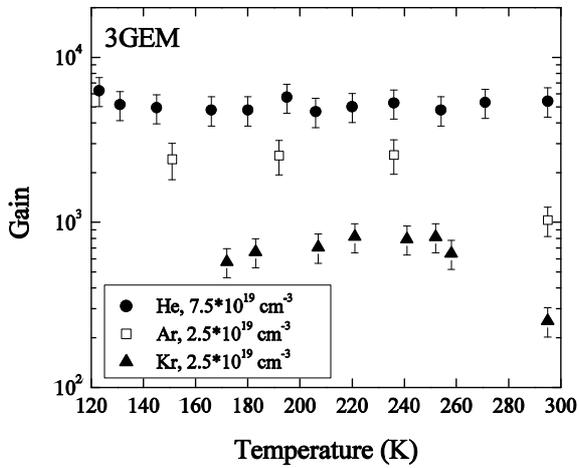

Fig.3 Triple-GEM gain as a function of the temperature at constant operation voltages and constant gas densities in He, Ar and Kr. The appropriate atomic densities are indicated.

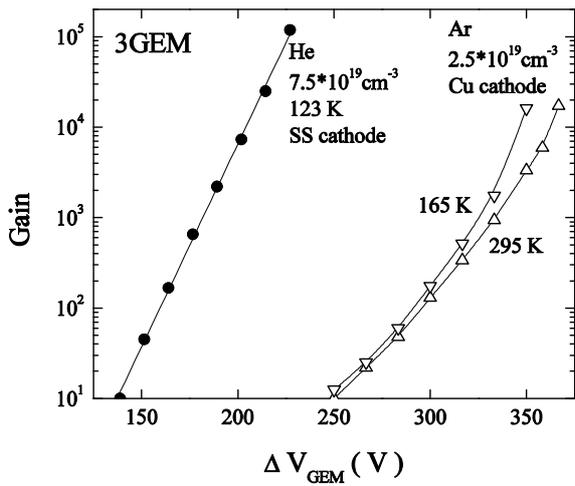

Fig.4 Triple-GEM gain as a function of the voltage across each GEM in He and Ar. The appropriate temperatures and atomic densities are indicated.

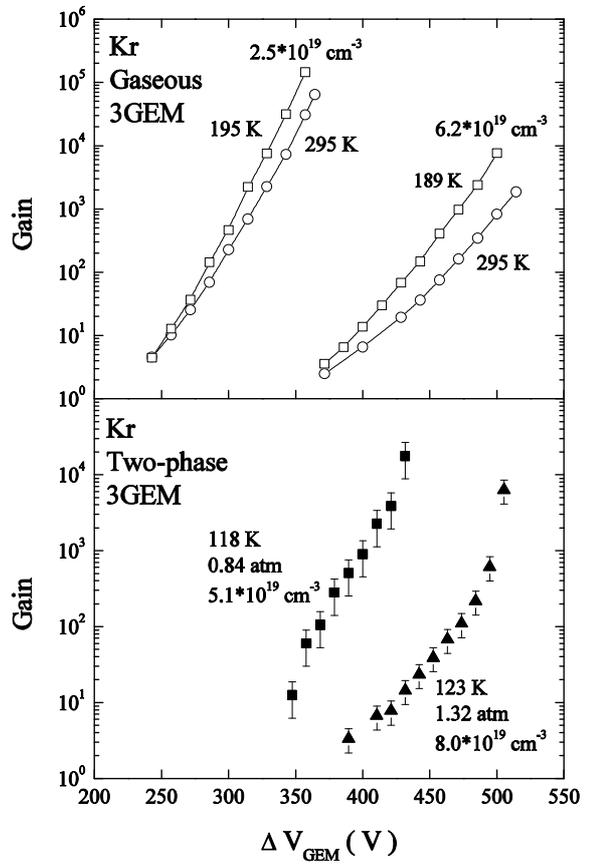

Fig.5 Triple-GEM gain as a function of the voltage across each GEM in Kr in the gaseous (top) and two-phase (bottom) modes. The appropriate temperatures and atomic densities of the gas phase are indicated.

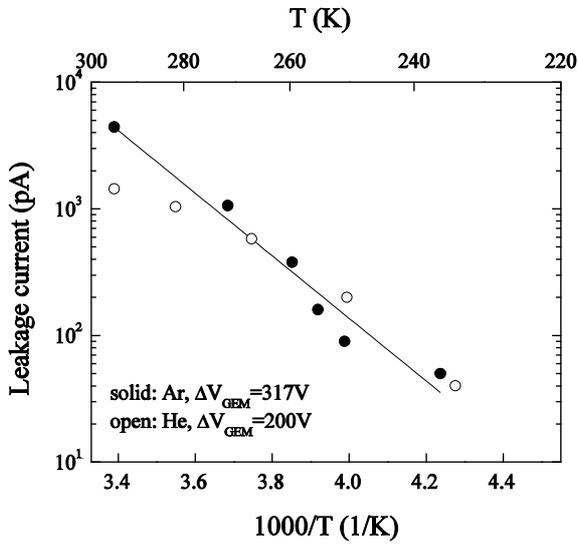

Fig.6 Leakage current of a single GEM as a function of the reciprocal temperature in He and Ar, at constant gas densities and constant GEM voltages. The atomic densities are $7.5\times10^{19}$ and $2.5\times10^{19}$ cm$^{-3}$, respectively.

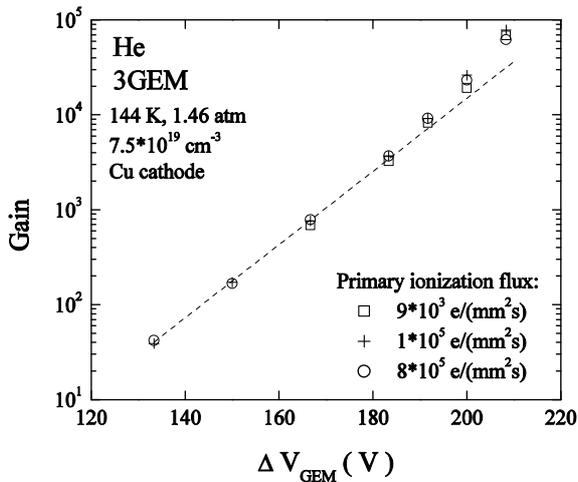

Fig.7 Triple-GEM gain as a function of the voltage across each GEM in He at 144 K, at three primary ionization fluxes induced by X-rays. The appropriate pressure and density are indicated.

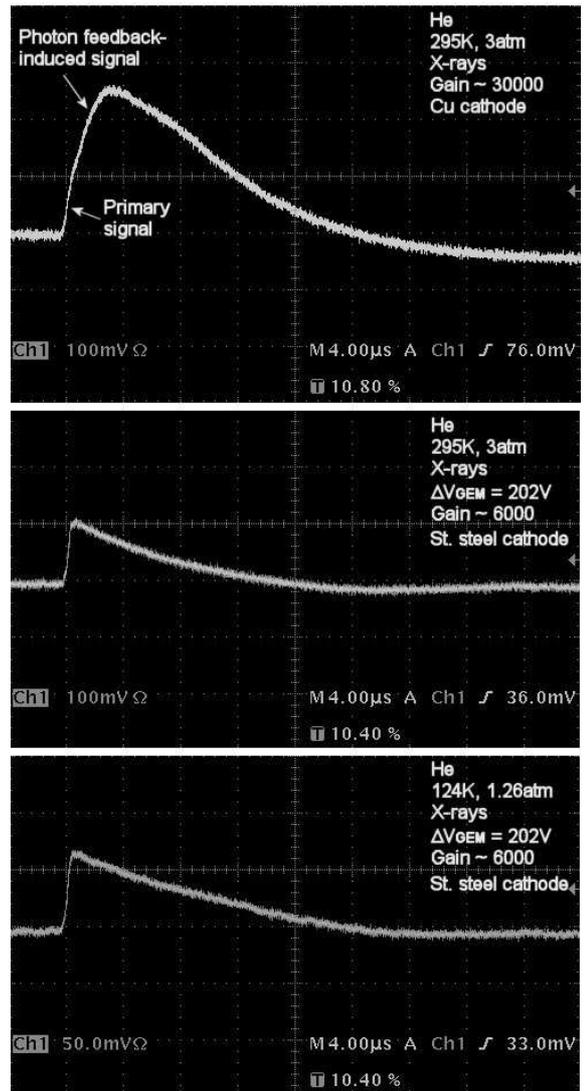

Fig.8 A typical anode signal from the triple-GEM in He, at room and low temperatures, at an atomic density of $7.5\times10^{19}$ cm$^{-3}$. Top: at 295 K and gain of $3\times10^4$, using the Cu cathode. Middle: at 295 K and gain of $6\times10^3$, using the stainless steel cathode. Bottom: at 124 K and gain of $6\times10^3$, using the stainless steel cathode. The signals are induced by X-rays.

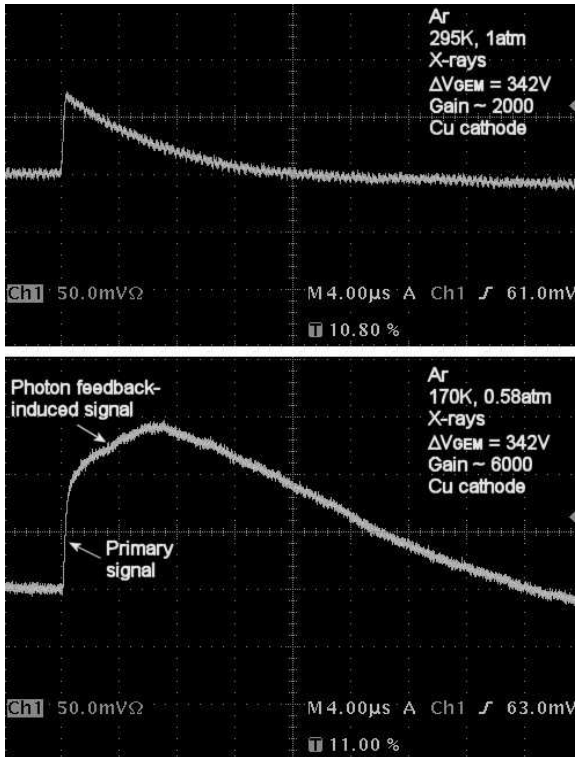
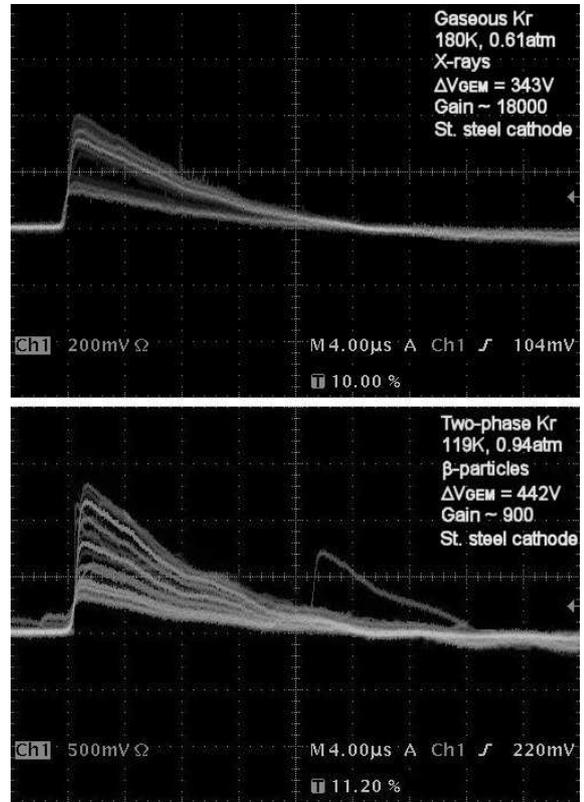

Fig.9 A typical anode signal from the triple-GEM in Ar, at room and low temperatures, using the Cu cathode, at an atomic density of $2.5\times10^{19}$ cm$^{-3}$. Top: at 295 K and gain of $2\times10^3$. Bottom: at 170 K and gain of $6\times10^3$. The signals are induced by X-rays.

Fig.10 Anode signals from the triple-GEM in Kr in the gaseous and two-phase modes. Top: in the gaseous mode at 180 K, 0.61 atm and gain of $1.8\times10^4$. Bottom: in the two-phase mode at 119 K, 0.94 atm and gain of $9\times10^2$. The signals are induced by X-rays and β-particles, respectively.

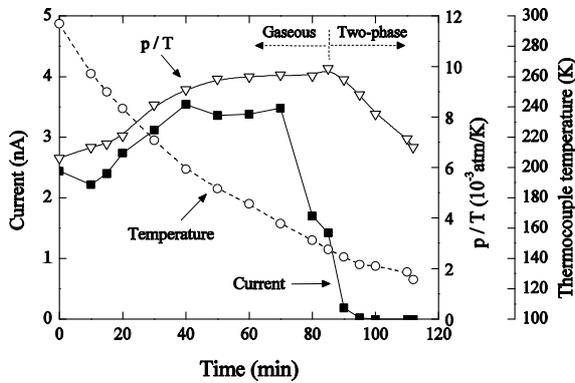

Fig.11 Time evolution of several parameters during the cooling cycle in the two-phase mode. Shown are the thermocouple temperature, pressure-to-temperature ratio and anode current recorded in the cathode gap at a field of 1.9 kV/cm (induced by X-rays).

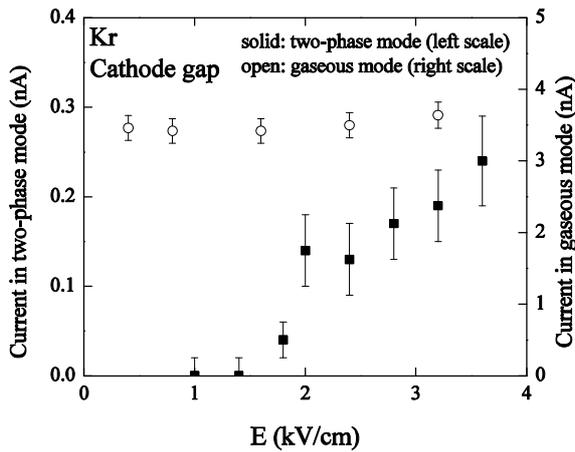

Fig.12 Anode current recorded in the cathode gap as a function of the electric field in Kr, in the two-phase (at 122 K and 1.23 atm) and gaseous (at 137 K and 1.67 atm) modes. The current is induced by X-rays.

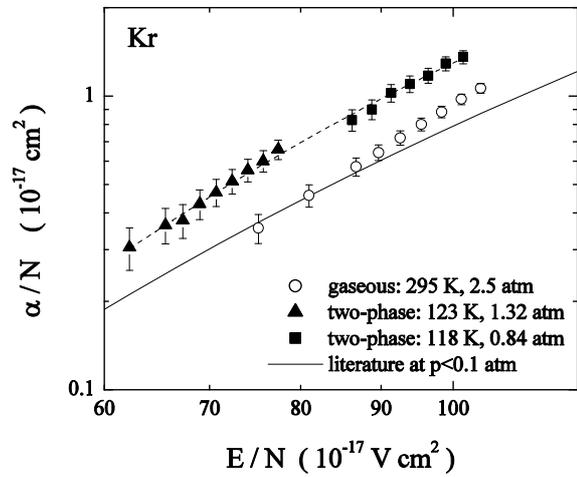

Fig.13 Reduced ionization coefficient as a function of the reduced electric field in two-phase and gaseous Kr, obtained from gain-voltage characteristics of Fig.5. Also shown are the data taken from literature obtained at low pressures [32] (solid curve) and the fit by two-phase data points (dashed curve).

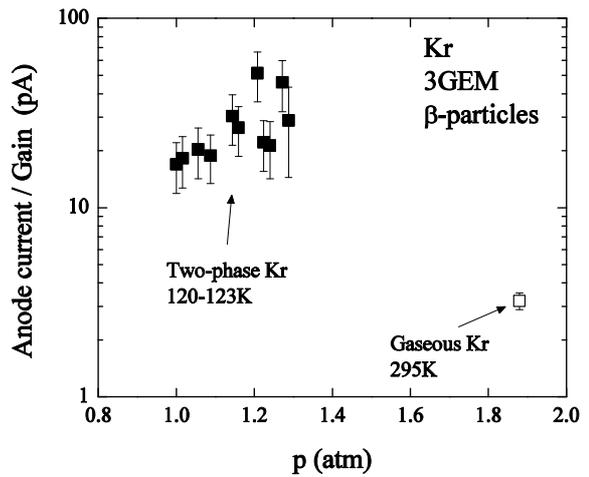

Fig.14 Ratio of the anode current to the gain of the triple-GEM as a function of the pressure, in gaseous and two-phase Kr, at gains varying in the range of $7 \times 10^2 - 1.5 \times 10^4$. The current is induced by β-particles.